\documentclass[prd,twocolumn,showpacs,nofootinbib,
preprintnumbers]{revtex4}

\usepackage{amsmath}
\usepackage{amsfonts}
\usepackage{graphicx}
\usepackage{dcolumn}

\def\be{\begin{equation}}
\def\ee{\end{equation}}
\def\ba{\begin{eqnarray}}
\def\ea{\end{eqnarray}}
\def\bs{\begin{subequations}}
\def\es{\end{subequations}}

\usepackage{color}

\begin{document}

\title{The transient and the late time attractor tachyon dark energy: Can we distinguish it from quintessence ?}

\author{Amna Ali\footnote{email:amnaalig@gmail.com }, M. Sami\footnote{email:sami.ctp@jmi.ac.in} and A.A. Sen\footnote{anjan.ctp@jmi.ac.in}}
\address{Centre of Theoretical Physics, Jamia Millia Islamia,
New Delhi-110025, India}

\begin{abstract}
The string inspired tachyon field can serve as a candidate of dark
energy. Its equation of state parameter $w$ varies from $0$ to $-1$.
In case of tachyon field potential $V(\phi)\to 0$ slower(faster)
than $1/\phi^2$ at infinity, dark energy(dark matter) is a late time
attractor. We investigate the tachyon dark energy models under the
assumption that $w$ is close to $-1$. We find that all the models
exhibit unique behavior around the present epoch which is exactly
same as that of the thawing quintessence.
\end{abstract}
\maketitle

\section{Introduction}
One of the most challenging problems of modern cosmology is
associated with late time acceleration of universe which is
supported by observations of complementary nature. According to the
standard lore, an exotic perfect barotropic fluid with large
negative pressure dubbed {\it  dark energy} can account for
repulsive effect causing acceleration\cite{review1,review2,review3}.
The simplest example of dark energy is provided by cosmological
constant $\Lambda$. The model is consistent with observation but is
plagued with difficult theoretical issues. The field theoretic
understanding of $\Lambda$ is far from being satisfactory and its
small numerical values gives rise to problems of {\it fine tuning}
and {\it coincidence}. A variety of scalar field models including
quintessence, tachyons, phantoms and K-essence has been investigated
in the recent years to address the problem\cite{review2,Paul,Kes}.
These models have some advantage over the cosmological constant: (i)
They can mimic cosmological constant at the present epoch and can
give rise to other observed values of the equation of state
parameter $w$ (recent data indicate that $w$ lies in a narrow strip
around $w=w_{\Lambda}=-1$ and is consistent with being below this
value). (ii) They can alleviate the fine tuning and coincidence
problems. 

The scalar field model, which is the simplest generalization of 
cosmological constant, is one with a linear potential \cite{linear}.
This model starts with a cosmological constant like behaviour where the
scalar field is frozen initially due to Hubble damping. Later on, it starts
rolling, but because the potential has no minimum, it leads to a collapsing 
universe in future. Hence universe in this model, has a finite history.

The more complicated scalar field models can broadly be classified into two
categories. Models in which scalar field mimics the background
(radiation/matter) being subdominant for most of the evolution
history. Only at late times it becomes dominant and accounts for the
late time acceleration. Such a solution is referred to as {\it
tracker}. In this case $w({\phi}) \simeq w_b$ ($w_b=0, 1/3$) before
the transition from matter like regime or {\it scaling regime} to
accelerated expansion. Tracker models are independent of initial
conditions used for field evolution but do require the tuning of the
slope of the scalar field potential. During the scaling regime the
field energy density is of the same order of magnitude as the
background energy density.

In second class of models, trackers are absent. Hence at early
times, the field gets locked ($w({\phi})=-1$) due to large Hubble
damping and waits for the matter energy density to become comparable
to field energy density which is made to happen at late times. The
field then begins to evolve towards larger values of $w({\phi})$
starting from $w({\phi})=-1$. In this case, for a viable cosmic
evolution, one chooses $\rho_{\phi} \sim \rho_{\Lambda}$ during the
locking regime which requires the tuning of initial conditions of
the field. The two classes of scalar fields are called Freezing and
Thawing models.

In case of standard scalar field (quintessence), there is a variety
of models which possess tracker solutions. In case of tachyon
field\cite{s1,as1} (motivated by string theory), there exists no
solution which can mimic scaling matter/radiation
regime\cite{samicop,samiothers,allthat,Paddy,staro,Bagla,AF,AL,GZ}.
These models necessarily belong to the class of thawing models.
Tachyon models do admit scaling solution in presence of a
hypothetical barotropic fluid with negative equation of state.
Tachyon fields can be classified by the asymptotic behavior of their
potentials for large values of the field: (i) $V(\phi) \to 0$ faster
than $1/\phi^2$ for $\phi \to \infty$. In this case dark matter like
solution is a late time attractor. Dark energy may arise in this
case as a transient phenomenon. (ii) $V(\phi) \to 0$ slower than
$1/\phi^2$ for $\phi \to \infty$ ; these models give rise to dark
energy as late time attractor. The two classes are separated by $
V(\phi)\sim 1/\phi^2$ which is scaling potential with
$w({\phi})=const$.

Since observationally, the equation of state parameter of dark
energy is very close to one, we can use this information to simplify
the dynamics. In case of thawing quintessence and phantom field, it
allows to obtain a generic expression for $w$ which represents the
entire class of quintessence and phantom models\cite{sen1,sen2}. In
this paper we apply the same technique to tachyon field which
belongs to the class of thawing models. With the current state of
observation, we address the issue of distinguishing the tachyon dark
energy from the case of quintessence.
\section{Dynamics of tachyon field}
In what follows we shall be interested in the cosmological dynamics
of tachyon field  which is specified by the Dirac-Born-Infeld
(DBI) type of action given by

\begin{equation}
\mathcal{S}=\int {
-V(\phi)\sqrt{1-\epsilon\partial^\mu\phi\partial_\mu\phi}}\sqrt{-g} d^4x
\label{Taction1}
\end{equation}
where on phenomenological grounds, we shall consider a wider class
of potentials satisfying the restriction that $V(\phi) \to 0$ as
$\phi \to \infty$. The parameter $\epsilon = \pm 1$ where the plus sign corresponds to the normal tachyon field which is non-phantom whereas with minus sign, one can model phantom type tachyon fields phenomenologically.
In FRW background, the pressure and energy
density of $\phi$ are given by
\begin{equation}
p_{\phi}=-V(\phi)\sqrt{1-\epsilon\dot{\phi}^2}
\end{equation}
\begin{equation}
\rho_{\phi}=\frac{V(\phi)}{\sqrt{1-\epsilon\dot{\phi}^2}}
\end{equation}
The equation of motion which follows from (\ref{Taction1}) is
\begin{equation}
\ddot{\phi}+3H\dot{\phi}(1-\epsilon\dot{\phi}^2)+\epsilon\frac{V'}{V}(1-\epsilon\dot{\phi}^2)=0
\end{equation}
where $H$ is the Hubble parameter
\begin{equation}
H^2=\frac{\rho_{\phi}+\rho_b}{3}
\end{equation}
The evolution equation can be cast in the following autonomous form
for the convenient use
\begin{eqnarray}
&& x'=-(1-\epsilon x^2)(3x-\sqrt{3}\epsilon\lambda y)\\
&& y'=\frac{y}{2}\left[-\sqrt{3} \lambda x
y-\frac{3(\gamma_{b}-\epsilon x^2)y^2}{\sqrt{1-\epsilon x^2}}+3\gamma_{b}\right] \\
&&\lambda'=-\sqrt{3}\lambda^2 xy(\Gamma-\frac{3}{2})
\end{eqnarray}
with
\begin{eqnarray}
 x=\dot{\phi},
 ~y=\frac{\sqrt{V(\phi)}}{\sqrt{3}H},~\lambda=-\frac{V_{\phi}}{V^{\frac{3}{2}}},~\Gamma=V\frac{V_{\phi\phi}}{V_{\phi}^2}
 \label{Gamma1}
\end{eqnarray}
where prime denotes the derivative with respect to $\ln(a)$. Here $\gamma_{b}$ is defined as $p_{b}=(\gamma_{b}-1)\rho_{b}$ for the background field. In our subsequent calculations, we shall assume a non-relativistic matter for our background field for which $\gamma_{b} =1 $.

An important remark on the autonomous system is in order. Let us
consider the inverse power law type potential $V(\phi)=V_0 \phi^{n}$
$(n<0)$. Eq.(\ref{Gamma1}) tells us that $\Gamma>3/2$ if $n<-2$
allowing $\lambda$ to increase monotonously for large values of the
field. In this case $\dot{\phi} \to 1$ or $w \to 0$ where as $w$
approaches the de-Sitter limit for $n>-2$ ($\Gamma<3/2$). These two
classes of tachyon potentials are separated by the inverse square
potential with constant $\lambda$ ($\Gamma=3/2$) which provides the
analog of scaling potential in case of tachyon. However, there is
major difference that in the present case, the field can only mimic
a hypothetical fluid with negative equation of state leading to
accelerated expansion. Unfortunately, the mass scale in the
potential turns out be larger than the Planck mass. The class of
potentials designated by $ -2<n<0$ is free from this problem and
gives rise to dark energy as late time attractor of dynamics.
\begin{figure}[t]
\includegraphics[width=80mm]{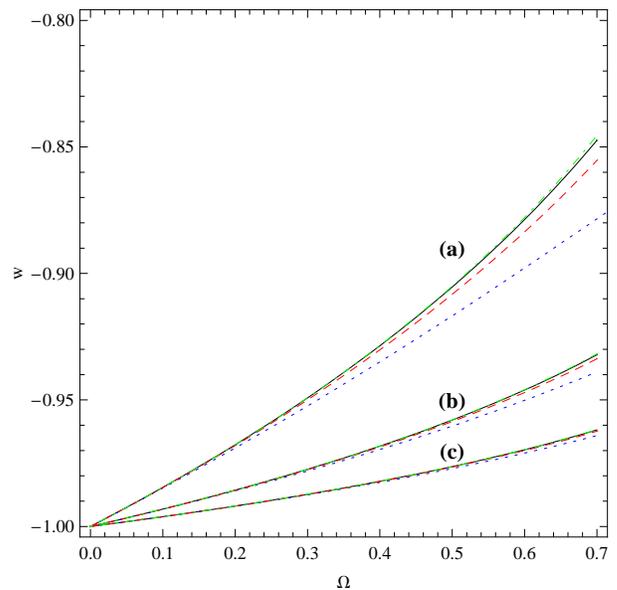}
\caption{Plot of dark energy equation of state parameter $w_{\phi}$
versus $\Omega_{\phi}$ for $0\leq \Omega_{\phi} \leq 0.7$ in case of
different values of $\lambda_{0}$ for nonphantom case i.e $\epsilon = 1$.
 The curves are for the potentials $V(\phi)=\phi^{-3}$ (dotdashed curve),$V(\phi)=\phi^{-2}$
 (dashed curve),$V(\phi)=\phi^{-1}$(dotted curve). The black solid line is for our analytical approximation (17). The sets (a), (b) and (c) are for $\lambda_{0} = 1, 2/3, 1/2$ respectively.}
\end{figure}
In the analysis to follow, it will be convenient to use the
following quantities
\begin{eqnarray}
\Omega_{\phi}=\frac{y^2}{\sqrt{1-\epsilon x^2}},~~\gamma_{\phi}=\epsilon(1+w)=\epsilon^{2}\dot{\phi}^2,
\end{eqnarray}
where $w= {p_{\phi}\over{\rho_{\phi}}}$ is the equation of state for the tachyon field. One can now express the autonomous equations through them:
\begin{equation}
\gamma_{\phi}'=-6\gamma_{\phi}(1-\epsilon\gamma_{\phi})+2\sqrt{3\gamma_{\phi}\Omega_{\phi}}\lambda(1-\epsilon\gamma_{\phi})^\frac{5}{4}
\end{equation}
\begin{equation}
\Omega_{\phi}'=3\Omega_{\phi}(1-\epsilon\gamma_{\phi})(1-\Omega_{\phi})
\end{equation}
\begin{equation}
\lambda'=-\epsilon\sqrt{3\gamma_{\phi}\Omega_{\phi}}\lambda^2(1-\epsilon\gamma_{\phi})^\frac{1}{4}(\Gamma-\frac{3}{2})
\end{equation}
The first two equations can be combined into one by a change of
variable from $a \to \Omega_{\phi}$
\begin{equation}
\frac{d\gamma_{\phi}}{d\Omega_{\phi}}=\frac{\gamma_{\phi}'}{\Omega_{\phi}'}=
\frac{-2\gamma_{\phi}(1-\epsilon\gamma_{\phi})}{\Omega_{\phi}(1-\Omega_{\phi})(1-\epsilon\gamma_{\phi})}+
\frac{2\sqrt{3\gamma_{\phi}\Omega_{\phi}}\lambda(1-\epsilon\gamma_{\phi})^\frac{5}{4}}{3\Omega_{\phi}(1-\Omega_{\phi})(1-\epsilon\gamma_{\phi})}
\label{Impeq}
\end{equation}
\subsection{Late time evolution}
From Eq.(\ref{Impeq}), one can see that for non-phantom and phantom
cases, i.e $\epsilon = \pm 1$, the equation is completely different
and hence one expects to have different evolutions for
$\gamma_{\phi}(\Omega_{\phi})$ for non-phantom and phantom cases.
\begin{figure}[t]
\includegraphics[width=80mm]{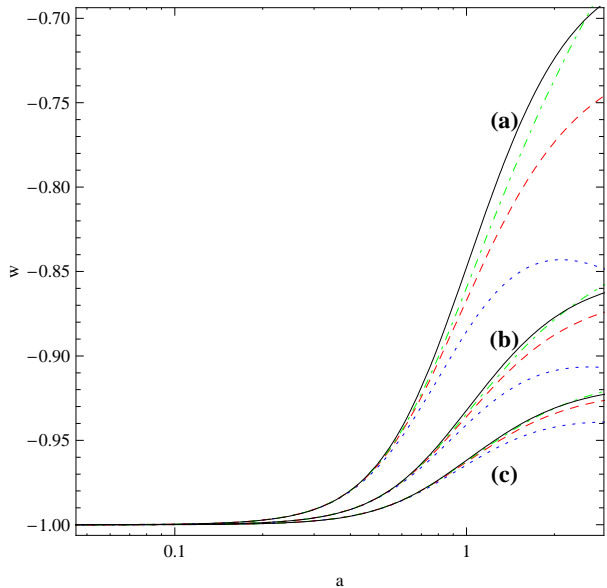}
\caption{Plot of dark energy equation of state parameter $w_{\phi}$
versus $a$ in case of different values of $\lambda_{0}$ for non-phantom case i.e $\epsilon = 1$.
 The curves are for the potentials $V(\phi)=\phi^{-3}$ (dotdashed curve),$V(\phi)=\phi^{-2}$
 (dashed curve),$V(\phi)=\phi^{-1}$(Dotted curve). The black solid line is for our analytical approximation (17) together with (20). The sets (a), (b) and (c) are for $\lambda_{0} = 1, 2/3, 1/2$ respectively.}
\end{figure}

But we are interested in the investigations of cosmological dynamics
around the present epoch where $\gamma_{\phi}<<1$. Secondly, in our
case $w({\phi})$ improves slightly beginning from the locking
regime, thereby, telling us that the slope of the potential does not
change appreciably. This implies that the potential is very flat
around the present epoch such that
\begin{eqnarray}
\frac{1}{V}\left(\frac{V_{,\phi}}{V}\right)^2<<1,~~\frac{V_{\phi\phi}}{V^2}<<1
\label{slowroll}
\end{eqnarray}
\begin{figure}[t]
\includegraphics[width=80mm]{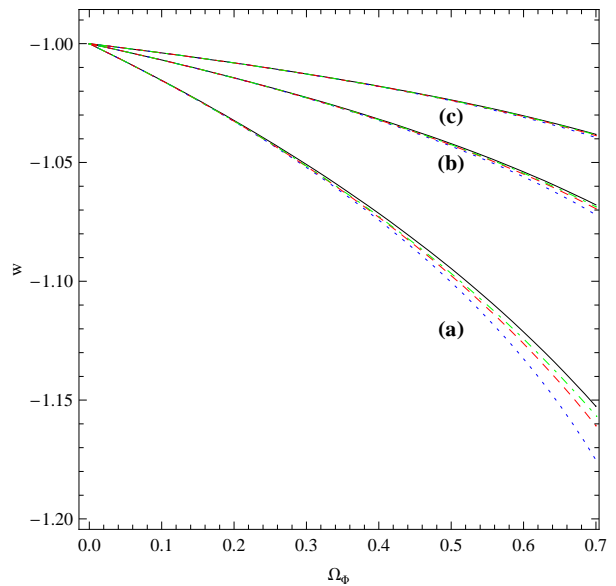}
\caption{Plot of dark energy equation of state parameter $w_{\phi}$
versus $\Omega_{\phi}$ for $0\leq \Omega_{\phi} \leq 0.7$ in case of
different values of $\lambda_{0}$ for phantom case i.e $\epsilon = -1$.
 The curves are for the potentials $V(\phi)=\phi^{-3}$ (dotdashed curve),$V(\phi)=\phi^{-2}$
 (dashed curve),$V(\phi)=\phi^{-1}$(dotted curve). The black solid line is for our analytical approximation (17). The sets (a), (b) and (c) are for $\lambda_{0} = 1, 2/3, 1/2$ respectively.}
\end{figure}
In case of field domination regime, the two conditions in
Eq.(\ref{slowroll}) define the slow roll parameters which allows to
neglect the $\ddot{\phi}$ term in equation of motion for $\phi$. In
the present context, unlike the case of inflation, the evolution of
field begins in the matter dominated regime and even today, the
contribution of matter is not negligible. The traditional slow roll
parameters can not be connected to the conditions on slope and
curvature of potential which essentially requires that Hubble
expansion is determined by the field energy density alone. Thus the
slow roll parameters are not that useful in case of late time
acceleration, though, Eq.(\ref{slowroll}) can still be helpful.
\begin{figure}[t]
\includegraphics[width=80mm]{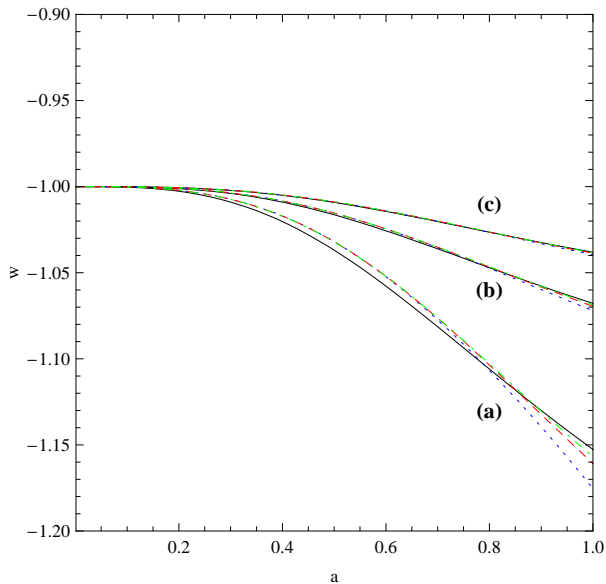}
\caption{Plot of dark energy equation of state parameter $w_{\phi}$
versus $a$ in case of different values of $\lambda_{0}$ for phantom case i.e $\epsilon = -1$.
 The curves are for the potentials $V(\phi)=\phi^{-3}$ (dotdashed curve),$V(\phi)=\phi^{-2}$
 (dashed curve),$V(\phi)=\phi^{-1}$(Dotted curve). The black solid line is for our analytical approximation (17) together with (20). The sets (a), (b) and (c) are for $\lambda_{0} = 1, 2/3, 1/2$ respectively.}
\end{figure}
 In view of the aforesaid, we can drop all the terms of order higher than  $\gamma_{\phi}$ in Eq.(\ref{Impeq})and
assume that the slope of the potential is constant,
$\lambda=\lambda_{0}$. These follow from the two slow-roll conditions (15) as we shall show later.  Evolution equation then simplifies to
\begin{equation}
\frac{d\gamma_{\phi}}{d\Omega_{\phi}}=\frac{-2\gamma_{\phi}}{\Omega_{\phi}(1-\Omega_{\phi})}+
\frac{2\lambda_{0}}{\sqrt{3}}\frac{\gamma_{\phi}^\frac{1}{2}}{(1-\Omega_{\phi})\sqrt{\Omega_{\phi}}}
\label{ImpeqL}
\end{equation}
Let us note that Eq.(\ref{ImpeqL}) is same as its counter part in
case of quintessence though the full Eq.(\ref{Impeq}) is different.
The difference between tachyon and quintessence dynamics is
represented by terms of higher order than $\gamma_{\phi}$. Thus if
we restrict our investigation of dark energy dynamics very close to
cosmological constant behavior, we can not distinguish tachyon dark
energy from quintessence. Also equation (16) is independent of
$\epsilon$. Hence $(1+w)$ for non-phantom case and $-(1+w)$ for
phantom case, have exactly similar evolution around cosmological
constant.


Eq.(\ref{ImpeqL}) can be transformed into a linear differential
equation with the change of variable $s^2=\gamma_{\phi}$,
 we have boundary condition $\gamma_{\phi}=0$ at $\Omega_{\phi} =0$.
 The resulting solution expressed in terms of $w(\phi)$\\
\begin{eqnarray}
1+w=\epsilon\frac{\lambda_{0}^2}{3}\left[\frac{1}{\sqrt{\Omega_{\phi}}}-(\frac{1}{\Omega_{\phi}}-1)tanh^{-1}\sqrt{\Omega_{\phi}}\right]^2\nonumber\\
=\epsilon\frac{\lambda_{0}^2}{3}\left[\frac{1}{\sqrt{\Omega_{\phi}}}-\frac{1}{2}(\frac{1}{\Omega_{\phi}}-1)ln(\frac{1+
\sqrt{\Omega_{\phi}}}{1-\sqrt{\Omega_{\phi}}})\right]^2 \label{wOmega}
\end{eqnarray}

Under the approximation $\gamma_{\phi}<<1$ which is justified about
the present epoch, all the tachyon models follow a general track
irrespective of the particular field potential. One can see from (17) that $1+w \sim O(\lambda^2)$. Hence the first slow roll condition ($\lambda <<1$) ensures that $1+w <<1$. We can quantify our second assumption that the slope of the potential does not change appreciably during the evolution as $\lambda^{'}/\lambda << 1$. Noting that $\gamma \sim \lambda^{2}$ and also $\gamma <<1$, one can then use eqn (13) to write
\begin{equation}
{V^{''}\over{V^{2}}} - {3\over{2}}{V^{'2}\over{V^{3}}} << 1;
\end{equation}
together with the first slow-roll condition, this ensures the second
slow-roll condition to be satisfied. We also show in figure 5, the
actual behavior of $\lambda$ for different potentials for
non-phantom case. This also shows $\lambda$ is constant during the
entire evolution for all practical purposes. One can also arrives
the same behavior for phantom case. In figure 1  and figure 3, we
show the our analytical approximation for $w(\Omega_{\phi})$ in
comparison with the numerical solutions of the exact equations for
different potentials with different initial values for $\lambda$ for
non-phantom and phantom cases. They show that our approximation
works reasonably well as long as $\lambda_{0}$ is small, i.e as long
as the slow-roll conditions are satisfied.

Next, we can use eqn (12) to solve for $\Omega_{\phi}(a)$ to
determine $w(a)$. assuming $\gamma_{\phi} <<1$, this gives
\begin{equation}
\Omega_{\phi} = \left[1+(\Omega_{\phi0}^{-1} -1)a^{-3}\right]^{-1}\label{Omega},
\end{equation}
where $\Omega_{\phi0}$ is the present day value of $\Omega_{\phi}$. Equation (17) and (19) gives the complete behavior for the equation of state $w(a)$ for tachyon fields with potentials satisfying the slow-roll conditions (15). One can also express the parameter $\lambda_{0}$ in terms of the present day value $w_{0}$ of the equation of state which is quite straightforward. This behaviours are shown in figure 2 and figure 4 for non-phantom and phantom cases.

 Similar to the case of
thawing quintessence, non-phantom tachyon models are restricted to a part of the
$w'-w$ plane. To specify the the limits, let us define a parameter
$X$

\begin{eqnarray}
X=-\frac{\ddot{\phi}}{H\dot{\phi}w}\nonumber =-\frac{w'}{2w(1+w)}
\to w'=-2Xw(1+w)
\end{eqnarray}
Since the Hubble parameter is determined by matter dominated regime
in the beginning of evolution, we find that $X=-3/2w\leq 3/2$
as $w\geq -1$ which leads to the upper limit, $w'<3(1+w)$.
The lower bound on $w'$ is estimated numerically (demanding that at
present $\Omega_{\phi} <=0.8$) as, $ w'>-.8(1+w)$ giving rise to the
permissible region of $w'$-$w$ plane
\begin{equation}
-0.8(1+w)<w'<3(1+w).
\end{equation}
\begin{figure}[t]
\includegraphics[width=80mm]{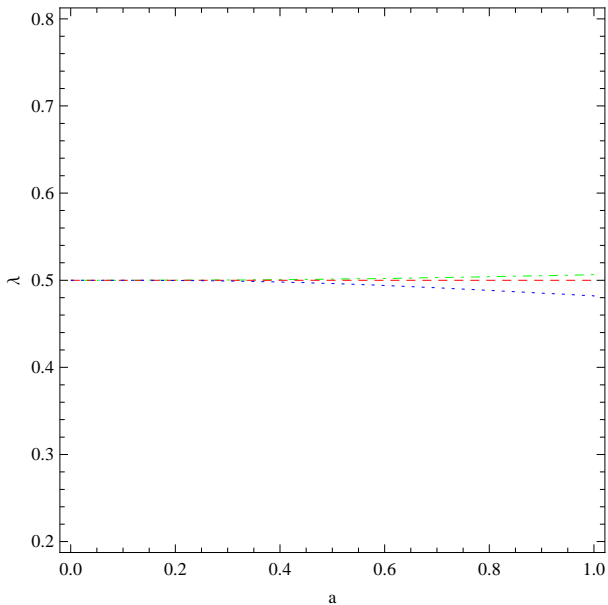}
\caption{Behavior of $\lambda$ as function of scale factor for
different potentials. We have chosen the initial value of
$\lambda_{i} = 0.5$. The Dotted, dashed and dotdashed curves
correspond to $V(\phi)=\phi^{-1},\phi^{-2},\phi^{-3}$ respectively.}
\end{figure}

In figure 3 we have shown this permissible region together with the
actual behavior for different potentials.

\begin{figure}[t]
\includegraphics[width=80mm]{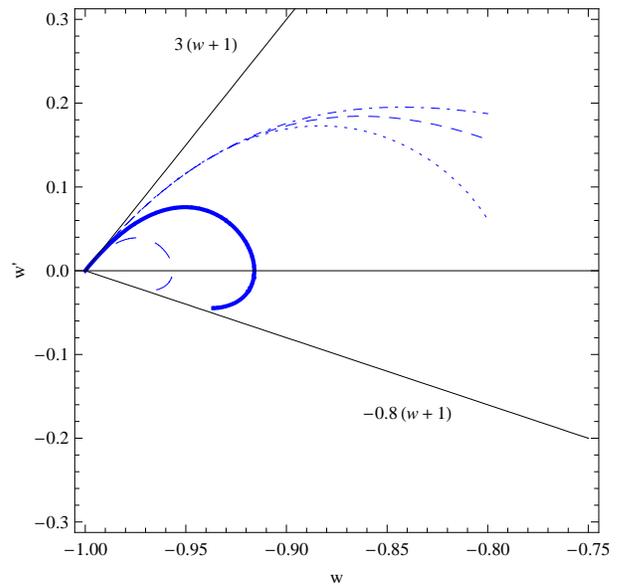}
\caption{The graph shows w-w' phase space occupied by the fields
. The
 upper bound and the lower bound correspond to $3(1+w)$ and $-0.8(1+w)$ respectively.
 The curves are for the potentials $V(\phi)=\phi^{-3}$ (dotdashed curve),$V(\phi)=\phi^{-2}$
 (Short dashed curve),$V(\phi)=\phi^{-1}$(Dotted curve),$V(\phi)=\phi^{-0.2}$(Thick curve),$V(\phi)=\phi^{-0.1}$(Long Dashed curve).The two thin lines are represent the upper and lower bound for the thawing models. The corresponding bounds are also specified.}
\end{figure}

\section{Observational Constraint}
The solution given by Eqs.(\ref{wOmega}) $\&$ (\ref{Omega}) for the
equation of state parameter $w$ versus the scale factor $a$ for
tachyon field under slow-roll conditions is exactly similar to that
for a canonical scalar field as obtained earlier in \cite{
sen1,sen2}. They have also constrained the two
parameters $w_{0}$ and $\Omega_{\phi0}$ of the model using the SNLS
(Supernva Legacy Survey)\cite{snls} and BAO data\cite{bao}. At present, we have the
Union08 compilation of the SnIa data which contains around 307 data
points\cite{kowalski}. This is world's published first
heterogeneous SN data set containing large sample of data from SNLS,
Essence survey, high redshift supernova data from Hubble Space
telescope as well as several small data sets. We use this data set
together with the BAO data from SDSS (Sloan Digital Sky Survey)\cite{bao}. The
$1\sigma$ and $2\sigma$ contour intervals for our model have been
shown in figure 4. From the figure, it is clear that one can not distinguish cosmological constant with a thawing dark energy models with present data although the phantom dark energy models are preferred.
\begin{figure}
\includegraphics[width=80mm]{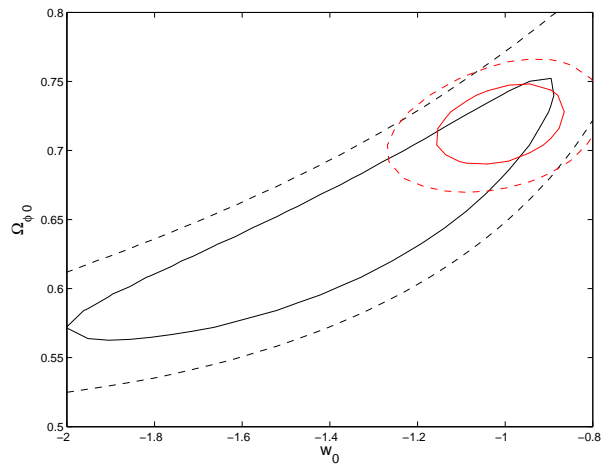}
\caption{Constraints in $w_{0}-\Omega_{\phi0}$ parameter space using Union08 compilation of SN data and BAO data. Black lines are for SN data only while Red lines are for SN+BAO data. Solid lines are for $1\sigma$ contour intervals while dashed lins are for $2\sigma$ contour intervals.}
\end{figure}

\section{Conclusions}
In this paper we have examined the DBI system with a
phenomenologically motivated class of run away potentials. In
general, the the tachyon dynamics crucially depends upon the
asymptotic behavior of the potential $V(\phi)$ at large values of
$\phi$. The inverse square potential gives rise to constant
equation of state which is determined by the slope of the potential,
$w=-1+\lambda^2/2$. We analysed the class of tachyon potentials with
dark energy and dark matter as late time attractors. Models in which
$V(\phi)$ decrease faster than $\phi^{-2}$ can give rise to
transient dark energy near the top of the potential and then mimic
dark matter as late time attractor. Since $\rho_{\phi}$ for tachyon
field scales slower than matter, its energy density for a viable
cosmic evolution should be fixed around $\rho_{\Lambda}$ at earlier
epochs allowing the field to freeze due to large Hubble damping.
Thus all the three classes of tachyon models belong to thawing type.
The data available at present allows to carry out investigations
around the present epoch with $\gamma_{\phi}<<1$. As soon as
$\rho_{\phi}$ becomes, comparable to matter density, field begins
evolve. The equation of state improves slightly starting from
$w(\phi)=-1$. Hence, the slope of the potential does not change
appreciably which we confirmed numerically. In the limit of small
adiabatic index of $\phi$ assuming $\lambda$ to be constant, we have
shown that the resulting evolution equations are same as in case of
quintessence which can be solved analytically. Our simulation shows
that the approximation is very close to the numerical results for
$\lambda<1$ around the present epoch. Deviations are possible in the
far future. We therefore conclude that tachyon dynamics is difficult
to distinguish from quintessence at least in the near future. We
also extended our analysis to the case of phantom tachyon. Again in
the region of interest, we find that phantom tachyon model is
difficult to distinguish from the ordinary phantom field. We also
constrained the parameters $w_{0}$ and $\Omega_{\phi0}$ for our
model using  the latest supernovae data along with baryon acoustic
oscillation BAO data. Our analysis shows some preference for phantom
energy.

The fact that all the scalar field dark energy models have a unique 
equation of state as long as they are in the 
slow-roll regime, makes a strong case for the  
$w(a)$ given by equations (17) and (19). It does not matter whether the 
scalar field has a canonical or non-canonical kinetic term. It is also the same 
for non-phantom or phantom scalar fields. We hope that this equation of state behaviour
for the dark energy will be considered seriously while fitting with the observational data coming from 
future experiments. 

\section{Acknowledgemnet}
AAS acknowledges the financial support provided by the University
Grants Commission, Govt. Of India, through the major research project
grant (Grant No:  33-28/2007(SR)).

\end{document}